# INFORMAL SOCIALIZATION IN PHYSICS TRAINING

Apriel K Hodari,[1] Shayna B Krammes[1]
Chanda Prescod-Weinstein,[2] Brian D Nord,[3] Jessica N Esquivel,[3] Kétévi A Assamagan[4]

[1]Eureka Scientific Inc, Oakland, CA 94602
[2]University of New Hampshire, Durham, NH 03824
[3]Fermi National Accelerator Laboratory, Batavia, IL 60510
[4]Brookhaven National Laboratory, Upton, NY 11973

## EXECUTIVE SUMMARY

This paper addresses issues related to the process of informal socialization into physics, particularly for senior graduate students and postdoctoral scholars. Many physicists' careers are built on the relationships they have and develop during these critical years.

***Socialization and Personal Agency***

The common conceptualization of mentoring is too broad to truly improve the experience and increase the success of novice scientists. Protégés will benefit much more if they take time to reflect and identify their needs before selecting a group of mentors, rather than relying solely on one person for support. This self-reflection also allows novice scientists to better understand what they are looking for, and perhaps also what to avoid so that they have the best experience possible.

***Pitfalls and Potholes***

Far too often, novice BIPOC scientists do not find a support network that is strong enough to counter the racism and isolation they face. There are little to no structures in place to prevent this—finding the right network is up to chance, and those who are not lucky often leave physics or STEM entirely. On the other hand, sometimes BIPOC women will find that they are being embraced by their physics community but for all the wrong reasons, and for only a fleeting time. The transition from pet to threat takes a toll both emotionally and professionally, and often leads to a difficult choice between career progression in a toxic environment, or starting over someplace safe.

***Lessons Learned***

Race matters in all settings, and claims of objectivity in physics are more of a dream than reality. Established physicists, particularly those in positions of power related to hiring and admissions, must understand how race functions in meritocracy, so that they may make more equitable decisions. The work continues as new hires and new students enter an institution and confront its culture. While a toxic culture and racism cannot be resolved overnight, faculty can begin to communicate nonverbal signs of value to their students and mentees immediately.

## INTRODUCTION

Research has analysed the experiences of many BIPOC as they navigated critical points in their scientific carers. For example, Jessica is featured in an analysis of career-life balance issues for STEM women of color (Kachchaf et al., 2015):

> *Jessica is a* [b]*lack woman with graduate degrees in a mathematics-based science. We interviewed her during her last year of a postdoctoral appointment at a private research university. At the time, she was single, job searching, and expressed great concern over her current and future prospects for achieving career–life balance. Jessica relayed numerous stressors she experienced in STEM, including a lack of understanding by her advisor about issues of balancing career and life, a lack of social connection with her lab mates, efforts to build and maintain a supportive peer network, and isolation.*
>
> *The unstated expectation* [was] *to eschew hobbies, rest, and even family … described by Jessica as "almost cartoon." She commented: "You should work all the time, this should be your prime, your main love is this work that you should do. And everything else should be secondary."*
>
> *Jessica had an advisor who heightened her anxiety about her future prospects for family and balance. … She repeatedly witnessed him prioritize work over his family. In one perhaps extreme example, Jessica's advisor chose to carry out routine data collection rather than attend the birth of his child. She described being "stunned" to hear him talk about this choice "as if it's an okay thing." Jessica recognized that her advisor's behavior was extreme, but experienced it as a disadvantage—she couldn't obtain support from her direct superior because of the disparity in their priorities.*

Jessica's story illustrates the intense pressure she felt to focus exclusively on her career at a time when she also felt she had a very limited window to have a family. The tension was amplified by the large gap between her priorities and those of her advisor.

Jessica's story highlights some of the dilemmas faced during the postdoctoral years, a critical career-building time for most physicists. For this paper, we are focused on the needs and struggles of senior graduate students and postdocs. While there are important issues to address in undergraduate STEM DEI, the *vast* majority of STEM DEI focuses on undergraduates. Senior graduate students and postdoctoral scholars are more vulnerable, in part because they are generally overlooked or seen as too complicated for most interventions.

The National Science Foundation has recognized the criticality of the postdoctoral years in building scientists' early careers. Because of this, any recipients of funding to support postdocs are required to provide specific mentoring plans to promote their success (NSF, 2021). Any proposal that does not meet the requirement is returned without review.

Given the criticality of these positions, we thought it was important to address topics that can derail progress. The goal here is to discuss behaviors and practices that undermine young scientists' success, and offer some insights into what advisors/supervisors, young physicists themselves, and organizations can do to counter them.

In this paper we present three sets of concepts—socialization and personal agency; barriers to success; and actions organizations can take to decrease/remove barriers. We feature both changes organizations can make and strategies young physicists can use to make progress despite the obstacles. Both are needed.

## Champions, Gangs, and Villages, Oh My

The most accepted definition of socialization is the process by which individuals move from being outsiders to insiders (Bullis & Bach, 1989) and mentoring relationships can facilitate this socialization process. However, the word "mentoring" is too broad, and means something different to almost everyone. Similarly, various forms of mentorship have historically been merged: their impact undifferentiated (Strayhorn & Saddler, 2009). As the term "mentoring" has such a broad definition, there should be a shift from person-based mentorship to identifying needs of a mentee/protege and how these needs can be met. A needs-based approach, like a formal research-focused mentoring relationship, has shown to be more helpful and useful to many people (Rockquemore, 2011).

Strayhorn and Saddler found that black undergraduate students engaging in a research-focused mentoring relationship with a faculty member, regardless of gender, have increased satisfaction with their college experience, which has been linked to retention (2009). While this study was published first, it is following what Kerry Ann Rockquemore would later describe as a necessary first step in successful mentoring: identifying a need (2011). Examples of needs that Rockquemore identified include emotional support, community, sponsorship, and a safe space and this range of needs underscores the importance of not relying on one person to fulfill all professional needs.

The findings of Strayhorn and Saddler's study show that a generalized, social type of mentoring is less useful or meaningful to black students (2009); professionals already in their chosen field echo this statement, according to Rockquemore (2011). Junior faculty require specific direction from their mentors, which they first must identify for themselves, in order to build a fulfilling mentor-protege relationship. While this process may have a flexible timeline, providing opportunities for physics <undergraduates> to seek out mentors, learn about their interests, and gain research experience will help keep them interested in physics and become more effective faculty once they reach that career stage.

### *Champions: The Importance of Sponsorship and of Using One's Social Capital*

> *Protege: one who is protected or trained or whose career is furthered by a person of experience, prominence, or influence*

One of the needs identified by Rockquemore is sponsorship. Research has shown that having a sponsor results in significant career gains, especially for historically marginalized communities (Center for Talent Innovation, 2019). Sponsorship focuses on influence and relationship and has been shown to have a significant impact on retention of historically marginalized populations. Sponsors leverage their influence, social capital, and relationships to advocate for their protege and rely on a three-way relationship between sponsors, proteges, and an audience to be swayed to the side of a sponsor's protege. Sponsorship also relies on amplifying and boosting a protege's accomplishments as well as facilitating connections and defending a protege's work and reputation (Chow, 2021).

In surveying sponsors, the identified needs from a protege were demonstrating trust, showing loyalty, and delivering 110% (Hewlett et al., 2011). As impactful as sponsorship is for marginalized groups, and despite high levels of ambition and drive, only 9% of black employees and 5% of Latinx employees have a sponsor compared to 13% of their white counterparts (Center for Talent Innovation et al., 2012). While a driver of a successful sponsor-protege relationship is trust, the pressure to conform to the status quo, and the racialized and gendered oppression that marginalized populations encounter, creates a breeding ground of distrust that negatively impacts this relationship building. The Center for Talent Innovation found that 26% of black leaders have the desire to sponsor employees that match their race, gender or ethnicity, compared with 7% of their white counterparts, yet because of toxic workplace culture and distrust, these potential sponsors are hesitating. These sponsors feel they don't have enough social capital to be an effective sponsor, and on the flip side, proteges are also hesitant to go into a multi-cultural sponsor-protege relationship due to potential disadvantages.

### *Gangs: The Simulated Dichotomy between Gatekeepers and Frenemies*

> *Gatekeeping: the activity of controlling, and usually limiting, general access to something*

Anchored in social network theory, which focuses on how socialization facilitates the transmission of information, Gaughan, Melkers, and Welch state that professional networks, instrumental networks in particular, share resources and information that enhance scholarly productivity (2018). Specific types of work-related social networks have been shown to be critical for career growth as they can impact access to tangible resources including sponsorship, support structures, and funding sources. However, research shows that historically marginalized scientists are often excluded or kept out of these professional networks that are tied to increased productivity, which in turn affects their professional success. Network size is related to networked social capital that potentially can be translated to increased productivity, and while BIPOC populations in STEM tend to have larger networks, their networks are composed of more advice resources compared to networks that are more instrumental or resource-based in nature. BIPOC populations seem to be further disadvantaged due to the seemingly lower publication productivity and impact in the field. Research further shows that advice-based networks have a consistent, direct, and negative impact on scholarly productivity. This contrast between advice-based networks and resource-based networks parallels the conversation above of needs-based sponsorship and relationship-based mentorship. This exclusionary practice of gatekeeping of resource-based networks due to race, ethnicity, and gender have shown to have a negative impact on BIPOC peoples career success, with intersecting identities suffering the most.

This toxic and exclusionary practice of gatekeeping resources from BIPOC populations has been shown to lead to the "crabs in a barrel effect" or anxiety related to perceived scarcity in resources. This in turn gives rise to the frenemy: a person who acts friendly despite a fundamental dislike or rivalry (Williams & Packer-Williams, 2019). In this context, the frenemy concept describes the relational aggressive behaviors women/girls engage in to advance their professional status. Williams and Williams take a critical race feminist stance to describe this concept through the intersectional lens of race, gender and class with a specific focus on how this frenemy concept can negatively impact the professional progress, and significantly affect the mental wellness of black women in particular. The competitive and egoistic culture of academia, coupled with the fact that academic socialization and access is buttressed by patriarchal whiteness, is in opposition to the cultural communal socialization of black women outside of academia. This academic paradox that pits black women against each other for access to power and professional capital, and removes the potential of collective power and communally celebrated triumphs, is a side-effect of the systemic issues that couple power and survival in academia to its proximity to whiteness.

### ***Villages: Counterspaces, Hyperspaces, and Peer Mentorship as an Act of Resistance***

> *Fictive Kinship: social ties that broaden mutual support networks, create community, and enhance social control beyond consanguineal and affinal ties*

Minnett, James-Gallaway, and Owens use a black femminist lens to study the benefits of a peer mentorship framework of black women doctoral students at predominantly white institutions and how this framework acts as a means to resist and protect black women doctoral students from the violence of white heteropatriarchy in the ivory towers (2019). Black women doctoral students face not only exclusionary practices, but antagonism traversing a toxic academic climate. Many search for more formalized faculty-student mentorship where one's race, gender, or class align, however, due to the underrepresentation of black women professors (Harley, 2008), black women professors become over-tapped which can lead to burnout and negatively impact the professor's career progression. Also, while a faculty-student pair may align demographically, as black people aren't monolithic, this doesn't guarantee a good fit. This peer mentorship framework acts not only as a support structure that is sorely missing for black women graduate students, it also acts to counter the cultural norms of individualism, competition, and meritocracy through collective power, mutual support, and unapologetic authenticity (Minnett et al., 2019). Due to the intentionality of a peer mentorship framework as an act of resistance, Minnett, James-Gallaway, and Owens focused on developing a framework focused on an informal faction that exists outside of institutional norms. This framework is also based on a horizontal and communal leadership structure, centers a deeper relationship building practice called sistership (James-Gallaway et al., forthcoming), and prioritizes holistic wellness over traditional forms of achievement. The horizontal leadership structure allows members' roles to shift from mentor to mentee and vice versa and is situated in black feminist socialization of communal and collective information sharing/gathering rather than the traditional academic socialization of competition and rigid power structures that lead to oppressive and suffocating mentor/mentee power dynamics. This peer mentorship framework is built on the basis of black women friendship, or sistership. The framework creates a space where black women graduate students can empower themselves, affirm each other's existence in the academy, and works to reverse, minimize, or heal potential mental health issues that arise from the oppression and dehumanization of black women in academia.

Ong, Smith and Ko discuss a broader safe space for women of color in STEM called counterspaces that aim to counter the cultural norms in STEM associated to white male scientists, including competitiveness, and individualism (2018). Ong et. al. detail five forms of counterpaces: peer-to-peer relationships, mentoring relationships, identity-based STEM conferences, student groups, and within STEM departments (Ong et al., 2018). Counterspaces work to combat feelings of isolation and the harm of microaggressions, and aid in the retention of women of color. Schipull, Quichocho, and Close also describe a counterspace for LGBTQ+ women and women of color (2020). The departmental practice of collective success of its practitioners with a social justice oriented pedagogy, places a counterspace at the center of a department rather than at the margins, allowing for LGBTQ+ women and women of color's physics identity and sense of belonging to thrive.

Evolving from counterspaces, created in the digital realm, leveraging social media, and materializing in the physical world when needed, comes the VanguardSTEM hyperspace (Isler et al., 2021). The VanguardSTEM hyperspace is an online platform developed through a black queer feminist lens that centers women, girls and gender non-conforming people of color in STEM. The goal of VanguardSTEM is to redefine STEM identity so that women and gender non-conforming people in STEM can unapologetically be fully without assimilation. The fluidity of a hyperspace is leveraged through VanguardSTEM's conference crashing and guerilla mentoring programs thereby creating a counterspace at conferences where women and non binary people of color often feel like outsiders. Lastly, the VanguardSTEM intersectional STEM framework is rooted in an intersectional STEM identity that redefines who and expands recognition of a STEM person.

Using recognition as a tool to increase STEM identity, and aligning with the principles of the CiteBlackWomen collective, VanguardSTEM developed a campaign to highlight women of color in STEM while still celebrating their cultural identity through the #WCWinSTEM, #QCWinSTEM, and the #GOCinSTEM hashtags (Smith, 2017). These hyperspaces are becoming more prevalent across STEM as many #BlackInX movements were created at the height of the civil unrest in the United States in 2020 including #BlackInPhysics, and #BlackInAstro (Brown et al., 2022; Grandison, 2021; Walker et al., 2022). These, built on the success of and theory behind Dr. Stephani Page's creation of #BlackandSTEM in 2014, which provided the basis for the idea that hashtags were a useful Twitter framework for helping black scientists not only find each other and build community but also discuss issues of significance to them.

***The Emperor has Blue Clothes***

With the motivations of the global uprisings following the murders of Ahmaud Arbery, Breonna Taylor, and George Floyd, many organizations embarked on or ramped up their DEI offerings to constituents or across their disciplines. Unfortunately, too many such organizations still failed to attend to troubling behaviors, policies and practices within their own walls, thus their outward-facing DEI activities amounted to little more than organizational gaslighting (NASEM, 2021). In fact, this is so commonplace, it's cliché.

Despite this, novices can employ strategies to increase their chances of succeeding (Hodari et al., 2016; Johnson et al., 2017; Ong et al., 2018). These include using their networks to avoid unsupportive supervisors and organizations; remembering their motivations for entering physics in the first place; learning to authoring multiple science identities as a skill set.

<u>Finding Sheep in Wolves' Clothing</u>. Access to networks is a major social determinant of success in academic and professional science. Those with access are in effect being voted "most likely to succeed." And this access is governed by people who already having it redistributing it. In other words, whether someone has access to these networks depends on who receives the necessary support in the form of not just encouragement, but also active and sustained mentorship. The distribution of mentoring and the access it sustains, as a resource, is uneven and tied to both the social and professional capital of the mentor and the potential mentee. Highly resourced students are more likely to attract mentorship and support from a highly resourced mentor. Lower resourced students may find themselves in environments where they are either disregarded or their sole source of support is a lower resourced mentor. For these students, "happenstance"—chance encounters between prospective mentees and prospective mentors who both take an interest in them and are in a position to support them—are of great import.

The significance of "happenstance" in student perseverance and success is both a symptom of a larger structural problem and a problem in and of itself. The way power is structurally distributed in society means that some students are essentially born with easy access to the networks and know-how that make the need for happenstance less significance because well-resourced communities ensure their members have access to those resources. Others will have an easier time experiencing happenstance because the barriers for them are lower: they look or sound like the archetype of a physicist and remind typical physicists of themselves—the chance encounter that underpins happenstance is more likely to occur because a person in a position of power is more likely to engage a new scholar. For the most marginalized groups, happenstance typically is a matter of luck and coincidence because the odds are stacked against them. Altogether, what we are describing is a totally dysfunctional system that helps students with social capital translate it into academic capital and makes it difficult for anyone else to manufacture academic capital and arrive at the resources necessary to succeed in both undergraduate and graduate programs.

In other words, happenstance is shaped strongly by the structural domain which governs how our institutions function, with implications for the interpersonal, cultural, and disciplining domains. These four domains come from Johnson's (2020) adaptation of Patricia Hill Collins's "domains-of-power" framework (Collins, 2009) to analyze the social and political dynamics of a physics department at a small predominantly undergraduate college. This framing provides a valuable analytic perspective through which we can study different aspects of how power works to shape the lives of scholars in physics. The interpersonal domain articulates how power is enacted and operates between individuals. This dynamic is in part governed by structural phenomena: what power each individual has to enforce their perspective, for example.

The cultural domain relates to ideas about cultural norms, e.g. "the ideal physicist." In Prescod-Weinstein (2020), the ideal physicist is mapped to Perry's "ideal patriarch" (2018), that is a person who is legally and socially understood to be a sovereign citizen by a society, which in the US is a straight, cis white man who, if he is disabled, is minimally disruptive to business as usual due to those disabilities. Finally, the disciplining domain focuses on rulemaking and rule enforcement. This refers both to who is in a position to make rules and who is most likely to be subjected to guidelines that restrict their participation due to their ascribed identities.

The apparent totality of these domains can look and feel intractable, especially to people who are not accustomed to struggling for access to needed resources. It is important for everyone to understand that each person has the capacity to have an impact on each of these four domains, particularly the interpersonal domain which is organized around our individual choices in relation to power we have access to. It is here that we can have an impact by working to ensure that the advantages granted by happenstance are not distributed according to social location but rather according to the needs of the students.

When considering the individual power that scholars have, it is important to be aware of one, novices' need to find the "happenstance", despite imperfect circumstances; and two, the power of one positive experience with a senior scientist, and how those who want to make a positive difference can act to ensure that they are the one. To address the first consideration, it is worth mentioning that happenstance should ultimately play a limited role in becoming a determinant of professional success. As long as it does, we face a scenario where developing a true meritocracy is impossible because "moving on up" depends extensively on luck. This is to say that it's necessary for departments to systematize equal access to the resources necessary for all student success, making happenstance less "do or die" for students and making valuable instances of happenstance more likely to occur (Johnson, 2020; Seymour & Hewitt, 1997; Thiry et al., 2019).

One form of happenstance that will likely remain important even after equal access to networks and mentoring is widely systematized and commonplace is the power of positive experiences with more senior scholars. Here "senior scholar" is used quite broadly. For an undergraduate, it might be a graduate student, postdoctoral researcher, or faculty member. Even one hour of positive, affirming contact has the potential to transform a student's sense of self in a physics space and enhance their perseverance, which is understood to be a key factor in keeping students in STEM majors (Johnson, 2020; Seymour & Hewitt, 1997; Thiry et al., 2019).

The challenge of course is to understand what constitutes a positive, affirming experience. It is not necessarily widely understood that being able to provide this to students is a skillset—and one that must be learned. Often, the way this is addressed is through assuming that people who share the students' identities ought to be responsible for supporting "diverse" students, which is to say, students who are not ideal physicists in the making. The assumption here has serious flaws: a person who shares a student's identity (e.g. race) may not have the same perspective on that idea and the social biases that shape their experiences. Moreover, even if they have a shared awareness of the issues, they may not have developed the skill that enables them to successfully talk to students about them in positive, productive ways.

Importantly, this is a skill that can be taught and learned, but few resources are available to specifically help minoritized/marginalized faculty articulate their own experiences, learn to address traumas associated with them, make them aware of the literature on these topics, and translate all of this into positive, affirming mentoring for students. Finally, the presumption is that faculty from traditionally marginalized groups will essentially take up the housework that ideal physicist faculty don't have to do. This leaves them with less time for their standard responsibilities and creates a disadvantage for those scholars when it comes to tenure and promotion. As the TEAM-UP report indicates in the context of African-American student success, "An equitable division of labor generally requires more than one faculty member of color in the department" (TEAM-UP, 2020). Organizations should ensure that minoritized mentors are being supported and recognized for their labor which makes it easier for them to say "yes" more frequently to students in need.

**Though it is true that a mentor who shares a minoritized students' ascribed identities is potentially uniquely situated to provide supportive mentoring, anyone who is willing to learn how can be a positive, affirming mentor**. In fact, it is important that all faculty engage in this kind of mentoring (TEAM-UP, 2020). This is a competency that should be seen as foundational to the expected competencies of faculty in physics, just like teaching introductory mechanics is something all instructional staff are expected to know how to do and do relatively-well. Though providing a positive mentoring experience to marginalized students is currently a specialty skillset, like teaching graduate particle physics, departments should be working toward expanding their faculty capacity to do exactly this kind of work. This baseline is necessary in order to achieve a learning and working environment that is equitable for both students and staff/faculty with mentoring responsibilities.

Effective mentoring has been extensively studied, which means that faculty and departments do not need to reinvent the wheel (NASEM, 2019). The Entering Mentoring curriculum described in NASEM 2019 identifies six behaviors from effective mentors: 1. align expectations; 2. assess understanding; 3. communicate effectively; 4. address equity and inclusion; 5. foster independence; and 6. promote professional development. In addition, the NASEM report identifies the importance of striking a balance between trust and privacy. It also highlights the significance of developing a mentoring toolkit, for example using individual development plans, mentoring compacts, etc.

Returning specifically to quality four, addressing equity and inclusion, we argue that this analytic should be used in developing a set of practices for the other five behaviors. Faculty should ask themselves about how expectations, communication, assessment, sensibilities about independence, and professional development are shaped by an understanding of students' ascribed identities (race, gender, etc) and specific community values. In the experience of the authors, a common refrain that students hear from advisors in physics is that maybe they should switch majors because their grades in another subject are higher. This is an example of a failure to adequately mentor and support successful professional development as a physicist, and it seems to disproportionately impact students of color. Mentorship should not involve stewarding students who want to be there out of the discipline, especially ones who represent an underrepresented demographic. There are other ways that faculty can end up being discouraging without intending to, for example through not acknowledging community values. The TEAM-UP report, for example, identifies that for African-American students, supporting their community is a major factor in their career decisions (TEAM-UP, 2020).

In relation, any mentoring should be culturally competent about features that typically play a role in the experiences of students from specific community backgrounds. At the same time, mentors have to be careful not to rely on stereotypes and to mentor the student in front of them. This requires being away of the diversity within communities. For example, as described in Imani Perry's *South to America: A Journey Below the Mason-Dixon Line to Understand the Soul of the Nation* (2022), black Southern experiences are distinct from northern, mid-west, and western experiences, and also they are regionally specific, e.g., Alabama and Florida are both southern but are also culturally distinct locales. Being attuned to the specifics of community dynamics can help a mentor be prepared to support the student in front of them.

Part of understanding the student in front of us is having an awareness that they may hold multiple marginalized identities. Not only do each of these identities come with distinct challenges under structural white supremacist patriarchy but also as described in intersectionality theory and the matrix of domination framework, they overlap to produce distinct experiences of disempowerment (Collins, 1990; hooks, 2000). Thus, a Native American two spirit/trans woman and a Native American cis man will have overlapping considerations under white supremacy but will also have distinct experiences due to the specific combination of white supremacy with (trans)misogyny. Both the man and the two spirit person will have different considerations from a white woman who is a member of the settler colonial majority. According to the LGBT+ Physicists Climate Survey, LGBT physicists who are also minoritized in another way experienced heightened levels of discrimination (2016)

Fostering independence for students from traditionally underrepresented/marginalized backgrounds also requires self-awareness and cultural competency on the part of the mentor. Language that might sound appropriate and supportive to the mentor can sound differently to someone who comes from a community that speaks a different English dialect, for example. Mentors have to be particularly attentive to stereotypes they may hold about students from certain groups, for example that people like them are poor, underprepared, and from communities that lack an interest in STEM. Some of these stereotypes are rooted in difficult realities—while there is no substantive evidence showing that any community has a lower interest in STEM, (queer and trans) people of color are more likely to be low-income and from lower resourced high schools than their white counterparts—but again it is important to mentor the student in front of you.

Recognizing the tendency to project these stereotypes onto students can help a mentor stop themselves from engaging in problematic behaviors and truly foster an environment that supports and sustains healthy development of students into independent scholars. This is particularly important to do in the case of marginalize students because the barriers they often face in academic spaces require them to be particularly well-prepared to be self-sufficient. As black youth often hear growing up, "You have to be twice as good to get half as far."

Part of fostering independence in young scholars is helping them learn to articulate what their needs are. What does a student need from a mentoring relationship? What kind of community support do they desire and would they find useful? A mentor does not have to be everything to a student, but they can help a student articulate the qualities of the village that they need to help them move through life successfully. It is important for mentors to get a sense of student needs and be ok with not being able to satisfy all of them. Instead, we should be ready to help students make the necessary connections and find the resources they need.

Mentors should also make sure to help students build strong professional networks where they can function as insiders in their discipline. This involves not only sending them to the typical professional society conferences and meetings on relevant topics, but also supporting mentee's participation in scientific affinity groups that support their communities. This can include connecting with identity-based organizations and groups such as: American Indian Science and Engineering Society (AISES), Society for the Advancement of Chicanos/Hispanics and Native Americans in the Sciences (SACNAS), the National Society of Black Physicists (NSBP), the National Society of Hispanic Physicists (NSHP), African-American Women in Physics Inc. (AAWIP), LGBT+ Physicists, Women in Science and Engineering (WiSE), &etc.

Rather than seeing these organizations as a time-consuming or distracting threat (as was the case for one of the author's former advisor), faculty should recognize these organizations can help sustain student interest in their academic work. It is also valuable—with permission of the scholar being asked to contribute labor—to connect students with scholars who share their identities and interests and may be able to provide them with community kinship.

Faculty should emphasize an open-door policy to all members of their group to avoid the appearance of playing favorites. This can feel uncomfortable, particularly when we don't identify with students. It may also be uncomfortable because this means students will come to us when they are experiencing difficult emotions. For example, students facing bias incidents may come to us with visible feelings of anger, underpinned by deep feelings of hurt. Here it is instructive to recall the words of black feminist poet and theorist Audre Lorde, "Anger is loaded with information and energy" (1997). Making room for students to feel their emotions in a safe environment where they are respected can help us maintain the open lines of communication necessary to teach students healthy approaches to angering professional situations. It also ensures that mentors are acknowledging legitimate responses to difficult situations, ensuring that students feel supported and like the door truly is open to them.

Importantly, while here we are emphasizing the experiences of students, these power dynamics do not dissipate upon conferral of the diploma. Postdoctoral researchers, faculty, and full-time lab staff continue to exist in an environment structured by the power dynamics of both the larger society and the unique features of physics as a professional space. Anyone in a position of seniority over others should be aware that these dynamics factor into the lives of people who are more junior to them, even if they are no longer students. Even those who have attained a high level of seniority and professional power still face racism, sexism, transphobia, ableism, and other forms of structural disempowerment.

There is an important connection to be drawn here between the experiences of marginalized students and more senior marginalized scholars. Though there is now a growing emphasis on supporting students, for example in the development of the American Physical Society National Mentoring Community and the publication of the TEAM-UP report, students are at risk when insufficient attention is paid to the considerations of people at the faculty level. The most obvious connection is that student feelings of isolation related to "being the only one" are heightened if the faculty are relatively homogeneous or at the least don't reflect the identities of people like them. This can affect their decision to stay in a physics major or a particular research area in physics. While there is overall an overproduction of PhDs relative to available faculty positions, this overproduction is not in the demographic groups traditionally labelled as "minorities." This means that each prospective PhD from these groups is precious and should be treated accordingly.

If marginalized scholars are forced out or choose not to stay in academia—or are not heavily supported in their efforts to build a career in academia—the downstream implications for mentoring and teaching are significant. Thus, providing effective and supportive environments for traditionally underrepresented group students requires doing the same for faculty who share those identities.

## NAVIGATING CAREER PROGRESSION

Aside from the difficulties in finding the right connections, novice physicists must navigate an array of obstacles as they build their careers. In this section, we will review some of these obstacles, and present navigation strategies others have found useful. We conclude with suggestions organizations can use to lower/eliminate these barriers.

### *When Happenstance Doesn't Happen*

Elaine Seymour and Nancy Hewitt investigated undergraduate persistence in STEM by contrasting the experiences of those that remain and those that leave (1997). Their primary finding was that the greatest difference between those that stay and those that leave is the serendipity of finding the right people and/or resources at critical moments, but no characteristic differences between the groups otherwise. Happenstance plays far too key a role in the experiences of STEM novices. It is both a point of commonality, and a discouragement that this experience happens regardless of demographics.

As has been documented in both the original study and its follow up (Seymour & Hewitt, 1997; Thiry et al., 2019), the consequences when this happenstance doesn't happen are disastrous. Inconsistent access to resources and support results in many students' and novice scientists leaving the discipline, taking their talents with them. That this inconsistency disproportionately pushes out minoritized people makes access to STEM participation serendipitous and deeply unjust.

### *Pet to Threat and other Dilemmas*

Psychologist Kecia M Thomas and colleagues defined a pet as a highly-credentialed "ethnic minority wom[a]n" who, upon entering new employment, is "welcomed into their workplaces, yet may be embraced for all of the wrong reasons" (2013, p. 276). Thomas' team summarized the experiences of five black women, from a larger study of faculty women across race, and found the socially assigned role of pet was characterized by five themes: tokenism, invisibility, nurtured/overprotected status, conformity/assimilation, and mistreatment.

In the book chapter where Thomas and colleagues presented their findings, they named and described the *pet to threat dilemma* professional women faced when they outgrew/rejected this role, thereby becoming a threat (Thomas et al., 2013). Themes associated with the threat role were challenging the status quo and microaggressions. In both cases, the woman's impressive qualifications, well-founded confidence, and high performance are the very things that proved threatening, and precipitated mistreatment, because they undermined their benefactor's comfort with the lie.

Hiring one/few minoritized people and treating them as pets, however well intentioned, amounts to a performance of progressive racial capital to impress other white people. As discussed in the *Power Dynamics* paper, such practices reinforce racial hierarchies, and do not ultimately benefit the professionals cast in this role, reducing them to little more than minority "Show and Tell" (Hodari et al., 2022; Reyes, 1988).

Targets of this role projection are acutely aware of the limits and costs, despite the veneer of acceptance. Whether targets outgrow their pet status, or manage to avoid it from the onset, the easily withdrawn acceptance leaves them isolated, and subject to the consequences of social isolation within a hierarchical power structure. For example, Juanita Johnson-Bailey described experiences where her positional power as the professor did not protect her from "academic bullying and incivility" from white men her students (2021). Professor Johnson-Bailey earned a promotion to Full Professor and an endowed distinguished professorship, yet one particular student "constantly and aggressively challenged" her "positional authority and [her] mastery". As Prof Johnson-Baily explained (2021):

> *The literature on* [b]*lack women's experiences in academia overwhelmingly posit that* [b]*lack women professors' positional power is trumped by the ability of students to activate a powerful system that has a vested interest in protecting its intended consumers, the students (Chepyator-Thomson, 2000; Johnson-Bailey & Lee, 2005; Sheared et al., 2010).* [Citations in the original.]

Johnson-Bailey's experience aligns Thomas' finding that black academic women cite the influence of white students positioning their black women faculty as pets, until their mastery and authority are highlighted by visible accomplishments, at which point they become threats and attract the hostility of the students (Myers, 2002; Reynolds-Dobbs et al., 2008; Thomas et al., 2013).

As DiAngelo often asks, "How does this [disturbing, repeated, disrespectful behavior] function" (DiAngelo & Menakem, 2020)? In Johnson-Bailey's case it downplayed the race, gender (and possibly class) advantage of the students, backed by institutional power. Despite the "delight" Johnson-Bailey took in having chosen "the career of my dreams … the Ph.D.s who I've minted, the courses I've created, and the publications that I've produced", her career was marred by "painful moments that left behind broken places" (Johnson-Bailey, 2021). Beyond the negative impact on her personally, these behaviors, and the colleagues and institutional practices that failed to confront them, reified racism, sexism, and classism as entrenched features of organizational culture, and reminded Johnson-Bailey that she would never be fully at home there, no matter what she accomplished.

These experiences are not unique to black women faculty. Catalyst reported that key aspects of the pet to threat dilemma are also experienced by women of color in accounting and law (Bagati, 2009; Giscombe, 2008). Further, people of color perceive these experiences very differently than their white colleagues (Donahue et al., 2021):

> [T]*he data show a largely failed understanding of the underrepresented minority experience on multiple levels, and how institutional racism has been widely accepted and inculcated into today's predominately white corporate settings. Even when people of color take the actions they are told will help them succeed in corporate America—getting a degree, getting a good job, working hard—43% of respondents were assumed to be in a service position, and 41% felt or were told they were only hired because of their race. This type of racism is so ingrained as 'normal', white coworkers remain unaware of how often people of color experience it, as well as the toll it takes on their careers and personal well being.*

The impact of these experiences often resulted in broken relationships and lost years of effort, because the inevitable disruptions that characterize the threat stage ruin personal connections required for good recommendation letters, particularly in communities where reputation is so critical to success. Even in the pet stage, the "benevolent treatment, which for many targets may initially feel good, simply maintains and reinforces the power and authority of the majority [white] group" (Thomas et al., 2013).

<u>Recommendations</u>. So what are the solutions to these dilemmas? What can people do to navigate this dilemma? Both Thomas and Stalling recommend seeking "safe spaces" outside of work both for psychological comfort and to build skills and experience in more productive environments (Stallings, 2020; Thomas et al., 2013). Stallings also advises building a "personal advisory board" comprised of mentors and peers, people a few levels above whose paths can be followed (Stallings, 2020).

Thomas suggestions additional strategies individuals can use, including reconceiving success markers, balancing the tensions between cultural pride and personal identity, and tapping into informal support structures (such as cohort friendships and leisure affinity groups) to build connections outside of the formal organizational hierarchy (Thomas et al., 2013).

Management Leadership for Tomorrow focuses on what organizations can do to guard against the damaging effects of pet to threat behaviors by making an organization-wide commitment to learning about BIPOC people's experiences; shifting to working in close proximity with diverse colleagues (over traditional DEI training); and tracking, frequently reviewing, and visualizing key data. Taken together, these strategies hope to decrease the damaging impact of this phenomenon, which undermines the many promising and high-achieving women.

<u>When Pet is Not Available</u>. As Thomas articulated, a key benefit of the pet role is access to collegiality and at least surface level acceptance. She highlighted that those who eschew pet status are thereby isolated. This is even more true for those for whom the pet role is simply unavailable, due to features of identity or appearance that are strongly undervalued by the institutional or societal culture: darker skin, LBTQIA+ identity, and body size (Harrison, 2010; Johnson, 2018; Lawrence, 2019).

Scholar Melissa Harris-Perry articulated that she benefitted from the privilege of being seen as a pet in a conversation with bell hooks (hooks & Harris-Perry, 2013):

> *Because I'm light-skinned, and cis*[-gender]*, and straight, and have a white parent, and have access to all kinds of privileges, from birth, my bet is that I have been seduced by power. I don't think that mine comes at the end of my penis, but my bet is that my proximity to whiteness has in fact allowed me over and over again, a level of racial naivety, and a willingness to believe that if I could just get the right white folks to give me cover, that it will be OK. And I think that has everything to do with being embodied in this body and not another*.

As Harris-Perry articulated, her privilege was secured in part by her proximity to whiteness. For those not born with the privileges she enumerates—light-skinned, cisgender, straight, half-white—the pet status, and its respite from noncollegial treatment, however problematic and fleeting, is simply never available.

***Which Roadblocks are Worth Dismantling*?**

In 2016, the property-rental company Airbnb finally decided to respond to the many stories of racist and discriminatory hosts. Since then, the company has reported several different measures to reduce discrimination in the booking process, with the most recent being announced in January of 2022 (Fitzpatrick, 2016; Romo, 2022). Of course, during the five years since Airbnb decided to acknowledge this problem, discrimination on the platform persisted. In 2017, an Asian American woman, Dyne Suh, was refused service upon check-in because of her race. Now-banned host Tami Barker justified her refusal to rent to Suh because "it's why we have Trump… and I will not allow this country to be told what to do by foreigners [sic]" (Solon, 2017). In addition to being banned from the platform, the California department of fair employment and housing ordered Barker to pay Suh $5,000 in damages and to complete a course in Asian American studies (Solon, 2017).

Since this incident in California, a variety of other incidents have arrisen and a number of company responses have been proposed. In 2019, Airbnb settled a lawsuit with three African American women who were discriminated against due to their race (Romo, 2022). Over two years later, the company is preparing to test how displaying initials rather than full names may reduce the amount of racial discrimination. A field study published by Harvard Business School in 2016 found that potential customers with "distinctively African-American names" are 16% less likely to be accepted by the host than those with "distinctively white names," despite having otherwise identical profiles (Edelman et al., 2016). With this information available well before either incident described above, it seems that the company should have acted sooner.

While it seems clear that Airbnb is acting because the company wishes to reduce discrimination within its community, the measures taken to protect customers from discrimination simultaneously protect bigoted property owners from profit loss. The same 2016 Harvard Business School field study found that "hosts who reject African-American guests are able to find a replacement guest only 35% of the time," which suggests that these hosts would in fact prefer to lose money than accommodate black guests (Edelman et al.). Certain roadblocks function well as warning signs of an unsafe space for BIPOC people, transgender people, or other marginalized groups. Why should members of these groups want to give their money to discriminatory hosts? Anonymizing the customer experience on Airbnb makes this more likely to happen; and it cannot be guaranteed that all will go well once the names and photos are eventually released to the property owner; recall that Suh was just minutes away from check-in when she was denied service.

Despite Airbnb's efforts, they have yet to solve the problem of hosts discriminating against customers of certain races. If these hosts want to purposefully lose profit because they cannot stand to share their space with a BIPOC person, why not let them lose? Some black travellers, for example, have created websites to list safe hosts (Diakite, 2022). Resources like this allow for BIPOC customers to show up as themselves, rather than take a chance on the increased anonymization that Airbnb is currently proposing. Alternatively, there are steps one can take to let bigots screen themselves out. Some BIPOC people looking to use Airbnb will find a property they like, message the host, and wait to see if the dates suddenly become unavailable. This method allows the user to gauge the responsiveness of the host and de-anonymizes the experience, which helps to ensure that they do not unknowingly do business with bigots, or end up in a hurtful, potentially dangerous situation. Sometimes it is better, or at least more convenient, to let the racism expose itself, take note and move on.

***A Note on Belonging to Yourself First***

Research has shown how successful BIPOC physicists enact various tactics to garner achievement despite the difficulties (Hodari et al., 2016; Johnson et al., 2011; Ong, 2005; Ong et al., 2018). Among these is how research participants center their own agency. As expressed my Maya Angelou, this includes believing in themselves, and having a willingness to stand alone (Angelou, 1973):

Bill Moyers: *What price have you paid for that freedom?*

Maya Angelou: *You are only free when you realize you belong no place—you belong every place—no place at all. The price is high. The reward is great.*

…

Bill Moyers: *Do you belong anywhere?*

Maya Angelou: *I haven't yet.*

Bill Moyers: *Do you belong to anyone?*

Maya Angelou: *More and more. I belong to myself. I am very proud of that. I am very concerned about how I look at Maya. I like Maya very much. I like the humor, and the courage, very much. And when I find myself acting in a way that doesn't please me, then I have to deal with that.*

Angelou's words, as quoted and analysed by Brown, center the need for BOPIC people to belong to themselves first (Angelou, 1973; Brown, 2017). This level of authenticity opens the door to several confidence-supporting non-verbal tools, including owning the space you occupy.

Further, true and productive connections to champions, peer support groups, and other support structures are best built on proteges' willingness to bring their authentic selves. Even when considering decisions about who to work with and where to work, being clear about their true needs helps them serve as their own best advisor. Ultimately, novices must develop internal belonging and put it to work for themselves.

*A bird sitting on a tree is never afraid of the branch breaking, because her trust is not on the branch but on its own wings. Always believe in yourself.*

Unknown

Across all of these phenomena, there are quite a number of things organizations do to help novices navigate their careers. These include:

- Make resources and support services as ubiquitous as possible;
- Help novices develop research habits that are portable across many groups and topics;
- Help novices understand and navigate the actual landscape of a research career by telling the untold rules, including information about non-academic careers;
- Engage in organization-wide learning about *in situ* experiences of BIPOC people;
- Focus on supporting opportunities aligned with novices' values and interests, rather than simply trying to reproduce traditional/existing trajectories;
- Track, review, and visualize key data; and
- Confront oppression directly when evidence suggests it is present; neutral policies are insufficient.

## FILLING POTHOLES AND LOWERING HURDLES

Choosing a graduate physics program is a big decision that comes with high financial stakes for most prospective students. Finding the right program, especially for a BIPOC person, is crucial not only to their success in their studies but also to their overall wellbeing. Considering how costly an investment earning any degree is, students do not want to settle for an institution in which they will not find support. Oftentimes, though, they have to settle, especially BIPOC students, LGBTQ+ students, and students who do not come from a financially privileged background.

The very first hurdle that any student encounters is the application. Institutions have the power to lower this hurdle and, in fact, many already have (*Graduate programs that don't require the physics GRE*). The American Astronomical Society first recommended that graduate physics programs stop requiring applicants to submit GRE and/or physics GRE scores over seven years ago (Urry, 2015). In light of this information, students looking to study physics today should make strategic decisions about where they apply, because the importance assigned to standardized testing says a lot about a given institution.

The reason why continued requirements to submit GRE and/or physics GRE scores raise a red flag is because many of these institutions implement cut-off scores, despite a warning against doing so from the makers of the tests (*GRE Guidelines for the Use of Scores*, 2022). Many studies and scholars have pointed out the relationship between race and gender and standardized test scores (Potvin et al., 2017; Prescod-Weinstein, 2018; Young et al., 2021, p. 2), meaning that they are not even reliable predictors of ability or future success. Of course, graduate applications must be reviewed somehow. Nicholas T. Young et al. published a study in October, 2021 which examined admissions at the University of Michigan over a three-year period. The study found that, by implementing a thorough rubric which considered more than test scores and GPA, the demographics of admitted graduate students became more diverse (Young et al., 2021). The authors of this study now recommend that other institutions consider this approach, as well. Rubric-based holistic review is not only a more equitable approach to admissions, it has the added benefit of being customizable. Institutions can set and weigh metrics in ways that make the most sense to them, their local context, and the populations they are trying to reach.

We cannot overstate the fact that admissions is only the beginning of the work to be done to make physics education more equitable. We live in a society that is inequitable by design, where microaggressions, bias, and racism seep into everyday interactions. Physics cannot exempt itself from this by simply claiming, however confidently, that it is a culture of no culture. Accepting diverse students into physics programs and hiring diverse physicists post-graduation means nothing if institutions do not have the ability to retain them. At every level, there is work to be done. The following section on meritocracy will illustrate how racism and bias have permeated education as a whole in the United States. Understanding the extent of the problem is necessary for any institution genuinely looking to become more equitable.

***The Illusion of Meritocracy***

This section begins by unpacking prabdeep singh kehal's article "Merit as Race Talk: The Ontological Myopia of Merit Knowledge" which argues that today's emphasis on merit, and by extension our claims of meritocracy in physics, is an extension of eugenics masked as something else. We then use DiAngelo's work to help illustrate how racism and white supremacy function within the history of merit. Finally, will explore the changes over the last 45 years that have allowed meritocracy to flourish while simultaneously thwarting opportunities for BIPOC students and professionals.

kehal recounts that it was during the popularity of the eugenics movement that "the universalization of every student 'having merit' was inserted into educational practice," and that merit was determined from results of intelligence testing (Dikötter, 1998; kehal, 2019, p. 21). The acknowledgement that every student has merit looks like a positive thing on the surface, but given that this merit (or how much merit) was assigned based on intelligence test scores, it quickly becomes problematic. Under this practice, merit allows for the ranking of students based on race because intelligence testing at the time was explicitly designed to place whites in the top percent tiles. At best, this application of merit was a way to avoid being held accountable for ranking students by race. Whether intentionally or not, the white folks who created this testing and ranking system gave themselves enough cover to last a century.

Although the idea of merit in education was developed before eugenics was discredited in the 1930s, perhaps educators and policymakers anticipated the impending need to change their language. As Adolf Hitler embraced and publicly applauded the pseudoscience of eugenics, it became increasingly less appealing to the United States, or any enemy of Nazi Germany for that matter (Markel, 2018). Note how whiteness is centered in this narrative; rather than examining the racial harm caused by eugenics and discrediting it for that reason, the true motivation for abandoning it was to create distance between democracy-loving USA and fascist Germany. Unfortunately, this abandonment was largely superficial, as the ideas behind eugenics seeped into how merit was defined and applied.

While Hitler did conceptualize the ideal white "Aryan race," in most cases, especially in a US context in the 21st century, whiteness has power in its invisibility. DiAngelo describes how whiteness and white supremacy function and thrive in her book *White Fragility*. Individualism is crucial to the functioning of whiteness. White people see themselves as free of race. When one white person looks upon another, there is an understanding that each of them are different, unique. Despite their shared racial identity, it is unlikely that either white person will make assumptions about the other based on race. White people also get to be "just people" because white is considered neutral, or the default (DiAngelo, 2021, p. 56). The fact that this paper calls out white paper by race may be uncomfortable to white readers, who are used to having their neutrality affirmed no matter where they are. For example, at the shopping mall, all items described as "flesh-tone" will be suited for white skin; white people can always easily find a place where they fit in racially, and easily leave the rare space in which they do not (DiAngelo, 2021, p. 53). On the other hand, plenty of generalizations and stereotypes exist about BIPOC people and are commonly talked about in both subtle and direct ways. These stereotypes, combined with white peoples' belief that white means neutral, create and uphold barriers for BIPOC people at every stage of life.

More than 80 years after the debunking of eugenics, merit and meritocracy are subject to more questioning but not yet enough to substantially change how they function in US higher education, especially in the sciences. The pressure to assimilate is perhaps most prevalent in the hard sciences, as the "culture of no culture" inherently upholds whiteness as the standard. kehal explains that assimilation was not the automatic or only choice, but rather one that bigots were forced to accept following the *Brown v. Board of Education* decision in 1954. This is not to say that the supporters of the *Brown* decision had limited their hopes to assimilation. In fact, when Justice Warren wrote for the majority, he specifically acknowledged that separate cannot be equal because it negatively impacts black students' sense of self-worth (Prescod-Weinstein, 2018). This rationale uniquely centers the victims of segregation, black students, and aspires to protect them according to an interpretation of the Equal Protections Clause that acknowledges the historical injustices against black Americans. Importantly, not only was *Brown* was the first Supreme Court case to interpret the Equal Protection Clause this way, the decision also "stood for quarter century as the gold standard of *how* to interpret" the clause (Prescod-Weinstein, 2018, emphasis added).

Unfortunately, a new gold standard arrived in 1978 following the Supreme Court's decision in *Regents of University of California v. Bakke*. Allan Bakke, a white, male NASA engineer and aspiring medical student, cried "reverse discrimination" and won, thereby limiting the scope of affirmative action and altering the interpretation of the Equal Protection Clause ("Bakke decision," n.d.; Prescod-Weinstein, 2018). *Bakke* relieved white policymakers, educators, and employers of having to confront the history of racism and white supremacy in the United States when making decisions. Much like how the shift from eugenics to merit served to avoid explicitly talking about race, when *Bakke* was decided, it marked the beginning of diversity-as-industry. In other words, there has since been a shift towards diversity which has proven to ignore or disregard the "social value of integration programs" in favor of "a deeply felt concern about properly measuring 'merit'" (Prescod-Weinstein, 2018). This implied message is that minoritized people cannot be meritorious; as kehal argues, merit is simply a rebranded colonial intelligence test, never intended to allow anyone but whites to succeed.

Understanding the history of race in the United States is imperative to making meaningful progress when it comes to any discipline or setting, and physics is no exception. Since the Supreme Court refused to uphold the *Brown* interpretation of the Equal Protection Clause, we must take it upon ourselves, upon physics as a discipline and as a community, to adopt this interpretation. The history of racism and white supremacy informs how black people are treated to this day, so pretending as if this history no longer matters is both insulting and harmful. It is for this reason that the diversity industry of today so terribly misses the mark. The diversity industry as we know it is trendy, and participating in training sessions or workshops make white people feel good about themselves rather than tackle the racial inequities that exist in society at large, and especially not the racial harm they inflict on a daily basis (DiAngelo, 2021). For more about this, see our accompanying white paper on *Power Dynamics* (Hodari et al., 2022).

In her review of Jonathan Kahn's book *Race on the Brain*, theoretical particle physicist Dr. Chanda Prescod-Weinstein clearly illustrates how the diversity industry ignores the real problems and causes harm. The hyper-focus on diversity in recent years is just a disguise for meritocracy and white fragility. This process mirrors how ideas derived from eugenics shifted into merit a century ago, advertised as something that will recognize individuals for what they can contribute when, in reality, whiteness continues to be upheld as most desirable. Promoting diversity may sound like a nice idea, but means nothing unless it accounts for implicit *and* explicit bias against BIPOC people, particularly in the hard sciences.

In a workplace that describes itself as having a "culture of no culture," such as physics, it is difficult to have any discussion of race because the dominant narrative says that race doesn't matter, only science. Even in spaces created to discuss race, such as a diversity training, black people are constantly told that their experiences of racism, bias, and discrimination are wrong (Prescod-Weinstein, 2018). In *Nice Racism*, Robin DiAngelo describes a particularly painful interaction that she witnessed in South Africa, when a mixed-race group of people voluntarily came together to hear her presentation on white fragility (2021):

> *Once the floor was opened to questions, the first person to take the mic was a white woman, who jumped up and began credentialing herself as she marched across the room… Several people, including the moderator, asked her to please sit down and get to her question, but she ignored them, striding around the room, microphone in hand, taking up both physical and psychological space and frustrating everyone, oblivious to the very direct social cues around her. She eventually got to her point, declaring triumphantly, 'We just need to see what is in one another's hearts!'*
>
> *When the white woman finally sat down, a Black woman stood up and expressed very deep pain and anger about the devastating racial inequality in South Africa. This should have been a sobering moment for white people to bear witness to a level of anger that is rarely safely expressed in mixed space. White people could have responded by demonstrating humility, holding back, and listening. Or they could have validated her experience by sharing the insights gained by her powerful testimony. Unfortunately, none of those things happened. The next person to speak was a white man who stood up, faced the woman, and proceeded to lecture her at length on the "answer" to racism ('personal responsibility'). He was followed by another white man who gave his version of this lecture, also pontificating at length ('personal relationships'). A white woman delivered the final lecture of the evening ('let's not point fingers'). None of these white people had questions for the speakers or demonstrated that they had heard—much less understood—a single point made in my talk, by the other panelists, or by Black audience members.*

While this particular example takes place in South Africa, situations almost identical to this one occur in the United States essentially every time that a mixed-race group is brought together to discuss race, diversity, or implicit bias. The histories of racial injustice in the US and South Africa are well-known and thought of in the collective imagination as especially brutal. However, the problem is not isolated to places such as these. Even in countries *not* known for racial injustice, or those with a small black population, white fragility finds a way to persist. For example, in the Netherlands, when the tradition of "Black Pete" is discussed, black Dutch people are routinely dismissed when they try to explain how this tradition is racist and harmful (Lagewaard, 2021, pp. 1577-1578). All of these scenarios are examples of testimonial injustice, which we write about in detail in the accompanying white paper on *Power Dynamics* (Hodari et al., 2022).

Testimonial injustice, the unwillingness of whites to listen to and believe the testimonies that BIPOC people give about racism and discrimination, is perhaps the primary reason that diversity trainings of today are so ineffective. Workplaces and organizations obsess over implicit bias, and often present implicit bias in a way that completely ignores race and context. This is problematic because BIPOC people continue to experience explicit bias, share these experiences, and then find that they are left out of discussions about diversity and equity time and time again (Prescod-Weinstein, 2018). To begin to solve this multifaceted problem, white people, people in power, DEI committees, must listen to BIPOC people and incorporate their suggestions so that diversity training actually addresses the pertinent issues.

For those wrestling with meritocracy in academia, Özlem Sensoy and DiAngelo created a guide that both explains how faculty populations across the board have remained so homogenous, white, and instructs hiring committees on how to make their hiring practices more equitable. Discourses of "fit" and "merit" are used to avoid explicitly naming race, but these discourses function to maintain the racial status quo of a given institution (Sensoy & DiAngelo, 2017, p. 573). For example, a BIPOC teacher with experience in "urban" or "majority minority" settings may be passed over due to concerns about hiring a teacher who can effectively teach "all" students (Sensoy & DiAngelo, 2017, p. 573). Why is this question not also asked of white candidates who only have experience teaching in majority white institutions? If "all" students of every race are to be adequately served, then faculty hiring committees must let go of this coded language and take their self-proclaimed commitments to diversity seriously.

While the meaning of diversity has been watered down and rendered unhelpful to BIPOC people, it does not have to be this way. Sensoy and DiAngelo explain how the "diversity question" is trivialized by most hiring committees. First, committees are not often required to ask diversity questions, or can manage to ask without addressing racial diversity, instead focusing on another aspect of identity. Because the diversity question is not treated as important by many institutions, candidates neglect to spend time to prepare answers; because the question is deemed unimportant, interviewers do not press candidates for meaningful responses or elaborations (Sensoy & DiAngelo, 2017, p. 570). To attempt to fix this, institutions must first have some kind of commitment to diversity and critical consciousness which hiring committees can use as benchmarks when assessing candidates. Rather than interpreting a candidate's experience teaching in a predominantly insert-race-here institution, the committee may ask which authors they taught in class, and judge that list based on its diversity (Sensoy & DiAngelo, 2017, p. 571). There must be a shift in which qualifications of a candidate are deemed important, and a conscious effort to interrupt whiteness and the maintaining of the status quo. The answers have been spelled out by many different groups, most importantly by BIPOC people, so while the work may be difficult, there is no excuse not to begin now.

***"Does Your Face Light Up when You See Them?" A Note of Seeing Protégés as Precious***

In a 2000 interview on The Oprah Winfrey Show, Nobel Laureate Toni Morrison described how she became intentional about showing her children that she cared about them, amid the bustle of challenging and hectic days as a single mother (Morrison, 2000). She said:

> *It's interesting to see, when a kid walks in the room—your child or anybody else's child—Does your face light up? That's what they're looking for. When my children used to walk in the room, when they were little, I looked at them to see if they buckled their trousers, if their hair was combed, or if their socks were up. And so you think that your affection and your deep love is on display, because you're caring for them. It's not. When they see you, they see the critical face. What's wrong now? But then, if you let your, as I tried, from then on, to let your face speak what's in your heart. Because when they walked in the room, I was glad to see them. It's just as small as that. You see?*

Here Morrison reminds us to communicate (nonverbally) as often as possible that our "children" are precious to us every day, even (especially) when they weren't virtuous or high achieving or perfect. It was a way of signaling value, no matter what else was happening.

As mentors, champions, and village elders to young scientists, we can take up this practice also. Minoritized members of our disciplinary communities often receive signals telling them that they do not belong. It costs us so little to counter those experiences with cues that tell them we are happy they are there. We can ensure that our face lights up when we see them.

In the context of an ongoing global pandemic, many people have lower capacity or desire to navigate the potholes and hurdles commonly associated with pursuing a degree or starting a new career. Institutions who remain hesitant or unwilling to help remove barriers that have, truthfully, always been problematic should make time to think deeply about why they believe these barriers are important. Chances are, the answer will be related to meritocracy.

The illusion of meritocracy was carefully crafted to center whiteness as ideal while constantly telling us that the rules were fair. It confidently asserts that the (white and male) people who find success and belonging in physics are in fact the only ones who worked hard enough to deserve it. Only for a brief moment in history did meritocracy face a challenge powerful enough to weaken it: *Brown vs. Board of Education*. Specifically, it was Justice Warren's interpretation of the Equal Protection Clause as something to be continuously pursued and infused into contemporary law and policy that interrupted the overwhelmingly white stream of new scientists, doctors, and college students.

When whites became tired of acknowledging the historic and systemic racism of the US, the current era of diversity was ushered in following the decision in *Regents of University of California v. Bakke.* Despite the testimony of BIPOC people from all walks of life, today's diversity programs continue to ignore the most pressing issues and serve as a way to make whites feel good about themselves; they want to feel as though they are "doing the work" needed to be "anti-racist" without actually challenging racism or white supremacy at all. This performative allyship is unacceptable.

Even within the toxic illusion of meritocracy which we all must navigate, concrete measures can be taken to lessen the harm done to BIPOC people. The solutions are easy to find, and the suggestions included in this white paper are just a small sample of what is available. As more students and professionals alike become aware of meritocracy's history and how it functions, they will expect answers and action. To ignore these expectations is to say that you do not care, that your institution does not care, and will surely push bright minds away towards people and places that do.

## Acknowledgements

The authors acknowledge the Heising-Simons Foundation Grant #2020-2374 which supported this work. We also thank all the non-author organizers for the 2021 A Rainbow of Dark Sectors conference: Regina Caputo, Djuna Croon, Nausheen Shah, and Tien-Tien Yu. We additionally thank Risa Wechsler for her organizational support.

## References

Angelou, M. (1973). *A Conversation with Maya Angelou* [Interview]. https://billmoyers.com/content/conversation-maya-angelou/

Atherton, T. J., Barthelemy, R. S., Deconinck, W., Falk, M. L., Garmon, S., Long, E., Plisch, M., Simmons, E. H., & Reeves, K. (2016). *LGBT Climate in Physics: Building an Inclusive Community*. A. P. Society. https://www.aps.org/programs/lgbt/

Bagati, D. (2009). *Women of Color in US Law Firms* (Women of Color in Professional Services Series, Issue.

Bakke decision. (n.d.). In *Encyclopedia Brittanica*.

Brown, C., Gonzales, E., Esquivel, j., Phillips, C., Sanders, V. A., Walker, A. L., Simpson, F., Edwards, M., Ramson, B., McAuley, O., Williams, G., & Platt, J. (2022). *Black In Physics*. https://www.blackinphysics.org/

https://twitter.com/BlackinPhysics

Bullis, C., & Bach, B. W. (1989). Socialization turning points: An examination of change in organizational identification. *Western Journal of Speech Communication*, *53*(3), 273-293. https://doi.org/10.1080/10570318909374307

Center for Talent Innovation. (2019). *The Sponsor Dividend*. https://coqual.org/reports/the-sponsor-dividend/

Center for Talent Innovation, Hewlett, S. A., Jackson, M., Close, E., & Emerson, C. (2012). *Vaulting the Color Bar: How Sponsorship Levers Multicultural Professionals into Leadership*. https://coqual.org/reports/vaulting-the-color-bar/

Chepyator-Thomson, R. (2000). Black women's experiences in teaching Euro-American students in higher education: Perspectives on knowledge production and reproduction, classroom discourse, and student resistance. *Journal of Research Association of Minority Children and Youth*, *4*(2), 9-20.

Chow, R. (2021). *Don't Just Mentor Women and People of Color. Sponsor Them.* Harvard Business Review. https://hbr.org/2021/06/dont-just-mentor-women-and-people-of-color-sponsor-them

Collins, P. H. (1990). Black feminist thought in the matrix of domination. *Black Feminist Thought: Knowledge, Consciousness, and the Politics of Empowerment*, *138*, 221-238.

Collins, P. H. (2009). *Another kind of public education: Race, schools, the media, and democratic possibilities*. Beacon Press.

Diakite, P. (2022, Feb. 15, 2022). *This Black Traveling Couple Just Created The Largest Black-Owned Airbnb List*. https://travelnoire.com/this-black-traveling-couple-just-created-the-largest-black-owned-airbnb-list

DiAngelo, R. (2021). *Nice Racism: How Progressive White People Perpetuate Racial Harm*. Beacon Press.
DiAngelo, R., & Menakem, R. (2020). *Towards a Framework for Repair* [Interview]. The On Being Project. https://onbeing.org/programs/robin-diangelo-and-resmaa-menakem-towards-a-framework-for-repair/
Dikötter, F. (1998). Race Culture: Recent Perspectives on the History of Eugenics. *The American Historical Review*, *103*(2), 467-478. https://doi.org/https://doi.org/10.2307/2649776
Donahue, K., Gilbert, T., Halpert, M., & Robertson Migas, P. (2021). *The Infuriating Journey from Pet to Threat: How Bias Undermines Black Women at Work*. MLT Blog. https://mlt.org/blog/the-infuriating-journey-from-pet-to-threat-how-bias-undermines-black-women-at-work/#:~:text=The%20%E2%80%9CPet%20to%20Threat%E2%80%9D%20phenomenon,that%20person%20had%20undermined%20them.
Edelman, B. G., Luca, M., & Svirsky, D. (2016, Jan. 25, 2016). *Racial Discrimination in the Sharing Economy: Evidence from a Field Experiment*. https://hbswk.hbs.edu/item/racial-discrimination-in-the-sharing-economy-evidence-from-a-field-experiment
Fitzpatrick, A. (2016, Sep. 8, 2016). *Airbnb CEO: 'Bias and Discrimination Have No Place' Here*. https://time.com/4484113/airbnb-ceo-brian-chesky-anti-discrimination-racism/
Gaughan, M., Melkers, J., & Welch, E. (2018). Differential Social Network Effects on Scholarly Productivity:An Intersectional Analysis. *Science, Technology, & Human Values*, *43*(3), 570-599. https://doi.org/10.1177/0162243917735900
Giscombe, K. (2008). *Women of Color in Accounting* (Women of Color in Professional Services Series, Issue.
*Graduate programs that don't require the physics GRE*. Washington University Physics Club. http://sps.wustl.edu/advice/applications/no-gre/
Grandison, T. (2021). *Black in X*. https://www.blackinx.org/
https://twitter.com/blackinxnetwork
*GRE Guidelines for the Use of Scores*. (2022). https://www.ets.org/gre/institutions/admissions/using_scores/guidelines?WT.ac=40361_owt19_180820
Harley, D. A. (2008). Maids of Academe: African American Women Faculty at Predominately White Institutions. *Journal of African American Studies*, *12*(1), 19-36. http://www.jstor.org/stable/41819156
Harrison, M. S. (2010). Colorism: The Often Un-Discussed-Ism in America's Workforce. *Jury Expert*, *22*, 67.
Hewlett, S. A., Marshall, M., & Sherbin, L. (2011). The Relationship You Need to Get Right. *Harvard Business Review*, 156. https://hbr.org/2011/10/the-relationship-you-need-to-get-right?registration=success
Hodari, A. K., Krammes, S. B., Prescod-Weinstein, C., Nord, B. D., Esquivel, J. N., & Assamagan, K. A. (2022). Power Dynamics in Physics. *arXiv*, Article arXiv:2203.11513 (physics.soc-ph). https://urldefense.com/v3/__http://arxiv.org/abs/2203.11513__;!!P4SdNyxKAPE!QvHMgtZnM6pbL0RjRd2LTTqxa2sJqlCbKmYfN5DFxo65MYI82fw40F9q9B0HDbo$
Hodari, A. K., Ong, M., Ko, L. T., & Smith, J. (2016). Enacting agency: The strategies of women of color in computing. *Computing in Science and Engineering*, *18*(3), 58-68.
hooks, b. (2000). *Feminist Theory: From Margin to Center* (2nd ed.). Pluto Press.

hooks, b., & Harris-Perry, M. (2013). *Black Female Voices: Who is Listening - A public dialogue between bell hooks + Melissa Harris-Perry*. The New School. https://www.youtube.com/watch?v=5OmgqXao1ng

Isler, J. C., Berryman, N. V., Harriot, A., Vilfranc, C. L., Carrington, L. J., & Lee, D. N. (2021). Defining the Flow—Using an Intersectional Scientific Methodology to Construct a VanguardSTEM Hyperspace. *Genealogy*, *5*(1), 8. https://www.mdpi.com/2313-5778/5/1/8

James-Gallaway, A., Griffin, A., & Kirkwood, M. (forthcoming). Engaging sistership: Toward a praxis of communal support amongst Black women graduate students. In D. Maye-Davis, T. B. Jones, & J. Andrews (Eds.), *Black sisterhoods: Black womyn's representations of sisterhood across the diaspora*. Demeter Press.

Johnson, A. (2020). An intersectional physics identity framework for studying physics settings. In A. T. D. Allison J Gonsalves (Ed.), *Physics Education and Gender* (pp. 53-80). Springer Nature.

Johnson, A., Brown, J., Carlone, H., & Cuevas, A. K. (2011). Authoring identity amidst the treacherous terrain of science: A multiracial feminist examination of the journeys of three women of color in science. *Journal of Research in Science Teaching*, *48*(4), 339-366. https://doi.org/10.1002/tea.20411

Johnson, A., Ong, M., Ko, L. T., Smith, J., & Hodari, A. (2017). Common challenges faced by women of color in physics, and actions faculty can take to minimize those challenges. *The Physics Teacher*, *55*.

Johnson, A. C. (2018). *Empowering faculty and students to create inclusive climates in STEM*. Georgetown University.

Johnson-Bailey, J. (2021). A scholarly journey to autoethnography: a way to understand, survive and resist. In V. Stead, C. Elliott, & S. Mavin (Eds.), *Handbook of Research Methods on Gender and Management*. Edward Elgar Publishing.

Johnson-Bailey, J., & Lee, M.-Y. (2005). Women of color inthe academy: Where's our authority in the classroom? *Feminist Teacher*, *15*(2), 111-122.

Kachchaf, R., Ko, L., Hodari, A., & Ong, M. (2015). Carer-life balance or women of color: Experiences in science and engineering academia. *Journal of Diversity in Higher Education*, *8*(3), 175-191.

kehal, p. s. (2019). Merit as Race Talk: The Ontological Myopia of Merit Knowledge.

Lagewaard, T. J. (2021). Epistemic injustice and deepened disagreement. *Philosophical Studies*, *178*, 1571-1592. https://doi.org/https://doi.org/10.1007/s11098-020-01496-x

Lawrence, K. A. (2019). *The Fat Agenda: An Analysis of Fatphobia, Race, Gender, Sexulaity and Black Womanhood* Georgia State University].

Lorde, A. (1997). The Uses of Anger. *Women's Studies Quarterly*, *25*(1/2), 278-285. http://www.jstor.org/stable/40005441

Markel, H. (2018). *Column: The false, racist theory of eugenics once ruled science. Let's never let that happen again*. PBS News Hour. https://www.pbs.org/newshour/nation/column-the-false-racist-theory-of-eugenics-once-ruled-science-lets-never-let-that-happen-again

Minnett, J. L., James-Gallaway, A. D., & Owens, D. R. (2019). Help A Sista Out: Black Women Doctoral Students' Use of Peer Mentorship as an Act of Resistance. *Mid-Western Educational Researcher*, *31*(2).

Morrison, T. (2000). *Does Your Face Light Up?* [Interview]. OWN youtube. https://www.youtube.com/watch?v=9Jw0Fu8nhOc

Myers, L. W. (2002). *A Broken Silence: Voices of African American Women in the Academy*. Greenwood Publishing Group.
NASEM. (2019). *The Science of Effective Mentorship in STEMM*. https://www.nap.edu/download/25568#
NSF. (2021). *Postdoctoral Researcher Mentoring Plan*. Retrieved from https://www.nsf.gov/pubs/policydocs/pappg22_1/pappg_2.jsp#IIC2j
Ong, M. (2005). Body projects of young women of color in physics: Intersections of gender, race, and science. *Social Problems*, *52*(4), 593-617.
Ong, M., Smith, J. M., & Ko, L. T. (2018). Counterspaces for women of color in STEM higher education: Marginal and central spaces for persistence and success. *Journal of Research in Science Teaching*, *55*(2), 206-245. https://doi.org/https://doi.org/10.1002/tea.21417
Perry, I. (2018). *Vexy Thing: On Gender and Liberation*. Duke University Press. https://doi.org/https://doi.org/10.1515/9781478002277
Perry, I. (2022). *South to America: A Journey Below the Mason-Dixon to Understand the Soul of a Nation*. HarperCollins.
Potvin, G., Chari, D., & Hodapp, T. (2017). Investigating approaches to diversity in a national survey of physics doctoral degree programs:
The graduate admissions landscape. *Physical Review Physics Education Research*, *13*(2).
https://doi.org/https://doi.org/10.1103/PhysRevPhysEducRes.13.020142
Prescod-Weinstein, C. (2018). *Diversity is a Dangerous Set-up*. https://medium.com/space-anthropology/diversity-is-a-dangerous-set-up-8cee942e7f22
Prescod-Weinstein, C. (2020). Making black women scientists under white empiricism: The racialization of epistemology in physics. *Signs: Hournal of Women in Culture and Society*, *45*(2).
Reyes, M. d. l. L. (1988). Racism in Academia: The old wolf revisited. *Harvard Educational Review*, *58*(3), 299-314.
Reynolds-Dobbs, W., Thomas, K. M., & Harrison, M. S. (2008). From Mammy to Superwoman:Images That Hinder Black Women's Career Development. *Journal of Career Development*, *35*(2), 129-150. https://doi.org/10.1177/0894845308325645
Rockquemore, K. A. (2011). *Don't Talk About Mentoring*. Inside Higher Ed. https://www.insidehighered.com/advice/2011/10/03/dont-talk-about-mentoring
Romo, V. (2022, Jan. 4, 2022). *Airbnb's new experiment to combat rental bias uses initials instead of names*. https://www.npr.org/2022/01/04/1070411574/airbnbs-new-experiment-racial-bias-discrimination-oregon
Schipull, E., Quichocho, X., & Close, E. (2020). *"Success Together": Physics departmental practices supporting LGBTQ+ women and women of color* Physics Education Research Conference,
Sensoy, Ö., & DiAngelo, R. (2017). "We Are All for Diversity, but...": How Faculty Hiring Committees Reproduce Whiteness and Practical Suggestions for How They Can Change. *Harvard Educational Review*, *87*(4), 557-580. https://doi.org/https://doi.org/10.17763/1943-5045-87.4.557
Seymour, E., & Hewitt, N. M. (1997). *Talking about leaving*. Westview Press.
Sheared, V., Johnson-Bailey, J., Peterson, E., III, S. A. C., & Brookfield, S. D. (2010). *The handbook of race and adult education: A resource for dialogue on racism*. John Wiley and Sons.
Smith, C. A. (2017). *Cite Black Women*. https://www.citeblackwomencollective.org/

Solon, O. (2017, Jul. 13, 2017). *Airbnb host who canceled reservation using racist comment must pay $5,000*. https://www.theguardian.com/technology/2017/jul/13/airbnb-california-racist-comment-penalty-asian-american

Stallings, E. (2020). When black women go from office pet to office threat. https://zora.medium.com/when-black-women-go-from-office-pet-to-office-threat-83bde710332e

Strayhorn, T. L., & Saddler, T. N. (2009). Gender Differences in the Influence of Faculty–Student Mentoring Relationships on Satisfaction with College among African Americans. *Journal of African American Studies*, *13*(4), 476-493. http://www.jstor.org/stable/41819227

TEAM-UP. (2020). *TheTime is Now: Systemic Changes to Increase African Americans with Bachelor's Degrees in Physics and Astronomy*. https://www.aip.org/sites/default/files/aipcorp/files/teamup-full-report.pdf

Thiry, H., Weston, T. J., Harper, R. P., Holland, D. G., Koch, A. K., Drake, B. M., Hunter, A.-B., & Seymour, E. (2019). *Talking About Leaving Revisited: Persistence, Relocation, and Loss in Undergraduate STEM Education* (E. Seymour & A.-B. Hunter, Eds.). Springer Nature Switzerland AG. https://doi.org/https://doi.org/10.1007/978-3-030-25304-2

Thomas, K. M., Johnson-Bailey, J., Phelps, R., Tran, N., & Johnson, L. (2013). Women of color at midcareer: Going from pet to threat. In L. Comas-Diaz & B. Green (Eds.), *The psychological health of women of color: Intersections, challenges, and opportunities* (pp. 275-286). ABC-CLIO. https://www.abc-clio.com/products/A3253C/

Urry, C. M. (2015). *President's Column: Rethinking the Role of the GRE*. American Astronomical Society. https://aas.org/posts/news/2015/12/presidents-column-rethinking-role-gre

Walker, A. L., Butler-Craig, N., Phillips, C., Polius, C., Muloma, K., Hadnott, B., Ivory, K., Zegeye, D., Joseph, T., & Tyler, D. (2022). *Black in Astro*. https://www.blackinastro.com/

Williams, W. S., & Packer-Williams, C. L. (2019). Frenemies in the Academy: Relational Aggression among Black Women Academicians. *The Qualitative Report*, *24*(8), 2009-2027.

Young, N. T., Tollefson, K., Zegers, R. G. T., & Caballero, M. D. (2021). Rubric-based holistic review: a promising route to equitable graduate admissions in physics. https://arxiv.org/abs/2110.04329